\documentclass{article}

\usepackage{arxiv}

\usepackage[utf8]{inputenc} % allow utf-8 input
\usepackage[T1]{fontenc}    % use 8-bit T1 fonts
\usepackage{hyperref}       % hyperlinks
\usepackage{url}            % simple URL typesetting
\usepackage{booktabs}       % professional-quality tables
\usepackage{amsfonts}       % blackboard math symbols
\usepackage{nicefrac}       % compact symbols for 1/2, etc.
\usepackage{microtype}      % microtypography
\usepackage{graphicx}
\usepackage{doi}
\usepackage{algorithm}
\usepackage{algorithmic}
\usepackage{graphicx}
\usepackage[font=small]{caption}
\usepackage{xcolor}
\usepackage{subcaption}

\usepackage{amsmath}
\usepackage{amsthm}
\usepackage{amssymb}
\newtheorem{definition}{Definition}
\newtheorem{theorem}{Theorem}
\newtheorem{corollary}{Corollary}
\newtheorem{proposition}{Proposition}
\newtheorem{assumption}{Assumption}

\usepackage[round]{natbib}

\title{Privacy Amplification by Subsampling in Time Domain}

\author{Tatsuki Koga\\
    Computer Science and Engineering\\
    University of California, San Diego\\
    La Jolla, CA 92093\\
    \texttt{tkoga@ucsd.edu}\\
    \And
    Casey Meehan\\
    Computer Science and Engineering\\
    University of California, San Diego\\
    La Jolla, CA 92093\\
    \texttt{cmeehan@eng.ucsd.edu}\\
    \And
    Kamalika Chaudhuri\\
    Computer Science and Engineering\\
    University of California, San Diego\\
    La Jolla, CA 92093\\
    \texttt{kamalika@cs.ucsd.edu}\\
}

\renewcommand{\shorttitle}{\textit{arXiv} Template}

\begin{document}
\maketitle

\begin{abstract}
Aggregate time-series data like traffic flow and site occupancy repeatedly sample statistics from a population across time. Such data can be profoundly useful for understanding trends within a given population, but also pose a significant privacy risk, potentially revealing \emph{e.g.,} who spends time where. Producing a private version of a time-series satisfying the standard definition of Differential Privacy (DP) is challenging due to the large influence a single participant can have on the sequence: if an individual can contribute to each time step, the amount of additive noise needed to satisfy privacy increases linearly with the number of time steps sampled. As such, if a signal spans a long duration or is oversampled, an excessive amount of noise must be added, drowning out underlying trends. However, in many applications an individual realistically \emph{cannot} participate at every time step. When this is the case, we observe that the influence of a single participant (sensitivity) can be reduced by subsampling and/or filtering in time, while still meeting privacy requirements. Using a novel analysis, we show this significant reduction in sensitivity and propose a corresponding class of privacy mechanisms. We demonstrate the utility benefits of these techniques empirically with real-world and synthetic time-series data. 
\end{abstract}

\section{Introduction}
\label{intro}
Time-series data about people's behavior can be sensitive, yet may contain information that is highly beneficial for society. As a concrete example, consider highway traffic data that is released by the California Department of Transportation~\footnote{https://pems.dot.ca.gov/} \citep{chen_freeway_2001}. The data contains aggregate statistics (number of cars, average speed, etc.) every five minutes gathered by physical sensors. 
Just like any other form of sensitive, aggregate data, this traffic data could expose the behavior of a single participant. In an extreme case, if a participant is the \emph{only} driver on a highway (say, late at night), then an adversary could track their movement through the entire road network. However, such data is absolutely vital for society to manage critical road infrastructure and strategize on where to augment it.
To this end, we aim to provide a privacy mechanism that releases a sanitized version of a given time-series while preserving its aggregate trends through time.

We adopt differential privacy (DP) \citep{dwork_calibrating_2006}, one of the most widely used privacy definitions, as our privacy measure. 
For data with bounded sensitivity, a first plausible DP mechanism is to add Gaussian \citep{dwork_algorithmic_2014} or Laplace \citep{dwork_calibrating_2006} noise proportional to the sensitivity. \citet{rastogi_differentially_2010} propose a second algorithm that adds noise to the first $k$ discrete Fourier transform (DFT) coefficients and ignores higher frequencies of the signal.
In both cases, the problem is that the sensitivity scales with $T$, the number of time steps. 
If an individual can theoretically participate at each timestep, then the worst-case influence they can have on the time-series grows linearly with $T$.
Thus, as $T$ becomes larger, the requisite noise variance increases, and the resulting output loses utility. In this paper, we take a new approach to addressing this longstanding problem using subsampling and filtering techniques from the signal processing domain. We find that random subsampling along with filtering can reduce sensitivity without significant losses of utility.  

Our starting point is to observe two key properties of real time-series data, which we exploit to design better mechanisms that add significantly less noise for a given $T$.
The first is that time-series data is often oversampled. 
Returning to our example, a transit analyst may be interested in observing how traffic changes on an hourly basis. This being the case, the five-minute sampling period of the sensor system is excessive and contributes very little to their utility. Indeed, with standard mechanisms, this excessive sampling rate may \emph{damage} their utility since it increases the number of time steps, $T$, many times beyond what is needed.
The second property is that a single person {\em{almost never}} contributes to the aggregate statistics at every time step.
It is implausible that an individual passes the same sensor on a highway every five minutes for more than a year. 

We formalize these two observations into mathematical assumptions,
and exploit them as follows. The first property motivates our use of subsampling, which trades these unnecessary extra time steps in exchange for reduced additive noise. The second property enables us to significantly boost this reduction in additive noise: we know that, feasibly, an individual can only participate in a limited number of time steps to begin with. This allows us to show that w.h.p. they will contribute to an even smaller number of timesteps \emph{after subsampling}, thus allowing us to significantly scale down noise added.

We additionally extend our results to a class of mechanisms that \emph{filter} and subsample the time-series. For the traffic analyst, the process of subsampling may be high variance if traffic changes significantly during an hour. In essence, which sample is chosen during that hour will have a large effect on the shape of the resulting time series. To mitigate this the analyst may wish to average the time series in time (filter) before subsampling. Unfortunately, this significantly complicates the privacy analysis used on subsampling alone. To address this we employ a novel concentration analysis that accounts for the effects of subsampling and filtering simultaneously.

This paper investigates how and in what circumstances subsampling and filtering enable us to achieve a better privacy-utility tradeoff for publishing aggregate time-series data under differential privacy.
Ultimately, by exploiting the two reasonable assumptions listed above, we make nontrivial gains in utility while maintaining a strong $(\epsilon, \delta)$-DP guarantee. 
Using a novel concentration analysis, we propose a class of filter/subsample mechanisms for publishing sanitized time-series data.
Our experiments indicate that---by using significantly less additive noise than baseline methods---our mechanisms achieve a better privacy-utility tradeoff in general for oversampled time-series.

\subsection{Related Work}
With regards to subsampling and Differential Privacy, prior work has shown that subsampling individuals (rows) helps amplify privacy \citep{kasiviswanathan_what_2011, beimel_bounds_2014, beimel_characterizing_2013, bassily_private_2014, abadi_deep_2016, balle_privacy_2018}; our setting is different from these results in that we consider subsampling time steps which are equivalent to columns or features, which has not been considered in prior work. 

The most closely related work in terms of publishing time-series data privately is by \citet{rastogi_differentially_2010}.
They propose an $\epsilon$-DP algorithm that perturbs only the first $k$ discrete Fourier transform (DFT) coefficients with the Laplace mechanism 
and zeros the remaining high frequency coefficients, since
adding noise to those results in the sanitized time-series being scattered. However, the sensitivity still scales with $T$ and the algorithm perturbs the first $k$ coefficient a lot if $T$ is large. In comparison, we provide an analysis using subsampling that introduces better scaling with $T$.
\citet{fan_real-time_2012} utilize the Kalman filter to correct the noisy signal by the Laplace mechanism and adopt adaptive sampling to reduce the published time steps, thereby reducing the number of compositions.
Their main focus is real-time data publishing, where we focus on offline data publishing.

Closer to our line of study is differentially private continual observation \citep{dwork_differential_2010, chan_private_2011}, which proposes a method to repeatedly publish aggregates  (in particular, count) under the continual observation of the data signal.
The main difference from our work is that they protect event-level privacy; namely, they try to hide the event that an individual participates in the time-series at a \emph{single} time step by perturbation. We hide participation in the \emph{entire} time-series, assuming that an individual can participate at multiple time steps. 

As for DP under filtering, \citet{ny_differentially_2014} investigate the DP mechanisms for general dynamic systems.
The work relates filters to an input with the sensitivity, but it does not address the issue that the sensitivity depends linearly on $T$.

More broadly, some work proposes new privacy definitions suitable for traffic (more generally, spatio-temporal) data \citep{xiao_protecting_2015, cao_quantifying_2017, cao_priste_2019, meehan_location_2021}.
Though these definitions are more appropriate for some situations, we believe the classic DP definition is the right choice for our setting, where there is a trusted central authority.

\section{Preliminaries and Problem Setting} \label{sec:prelim}
Among aggregate time-series data, we focus on a \textbf{count} data throughout this paper. 
More formally, we consider time-series signal $x \in \mathbb{N}^{T}$ of length $T$, where $x_{t}$ corresponds to the count of individuals' participation at time $t$.
The count signal $x$ is an output of a function $\mathrm{count}: \mathcal{X} \to \mathbb{N}^{T}$, where $\mathcal{X}$ is the input space.
We aim to publish the randomized version of $x$, denoting as $z$, with a privacy guarantee.

The appropriate privacy notion for our setting is differential privacy.
We say two datasets $D, D^\prime \in \mathcal{X}$ are neighboring if they differ on at most an individual's participation.
\begin{definition}[$(\epsilon, \delta)$-DP \citep{dwork_our_2006}]
A randomized algorithm $M$ satisfies $(\epsilon, \delta)$-DP if for any neighboring datasets $D,D^{\prime}$ and for any $S\subseteq \mathrm{range}(M)$,
\begin{align*}
    \Pr \{M(D) \in S\} \leq \exp (\epsilon) \Pr \{M(D^\prime) \in S\} + \delta.
\end{align*}
\end{definition}
One of the most famous DP algorithms is the Gaussian mechanism, which adds the Gaussian noise with a variance proportional to the $L_2$-sensitivity \citep{dwork_calibrating_2006}. 
The $L_2$-sensitivity of a function $f$ is the maximum difference between the outputs of $f$ from any neighboring datasets in terms of $\ell_2$-norm.
\begin{definition}
The $L_{2}$-sensitivity of a function $f:\mathcal{X}\to \mathbb{R}^d$ is 
\begin{align*}
    \Delta_{2,f} = \max_{D,D^\prime} \|f(D) -f(D^\prime)\|_{2},
\end{align*}
where $D$ and $D^\prime$ are neighboring datasets and $\|\cdot\|_{2}$ is $\ell_{2}$-norm.
\end{definition}
\begin{theorem}[Gaussian mechanism \citep{dwork_algorithmic_2014}]
Let $\epsilon \in (0,1)$ be arbitrary and $f:\mathcal{X} \to \mathbb{R}^d$ be a function. 
Then, for $c^2 > 2\ln(1.25/\delta)$, the algorithm $M$:
\begin{align*}
    M(D) = f(D) + (\xi_1, \ldots, \xi_{d})^{\top}
\end{align*}
satisifies $(\epsilon, \delta)$-DP,
where $\xi_i$'s are drawn i.i.d. from $\mathcal{N}(0, \sigma^2)$ and $\sigma \geq \frac{c\Delta_{2,f}}{\epsilon}$.
\end{theorem}
As we see from the Gaussian mechanism, the larger $L_2$-sensitivity leads to the higher noise variance.
Thus, we consider filtering and subsampling the input signal to reduce the sensitivity.

Filtering the signal is a well-used technique in signal processing. 
In particular, low-pass filters attenuate the effect of high frequencies and make the signal smoother.
Let $h \in \mathbb{R}^{T}$ be a filtering vector and $y \in \mathbb{R}^{T}$ be a filtered signal.
We express the filtering operation as
$y_{t} = \sum_{k=0}^{T-1} x_{k} h_{(t-k) \bmod T}$
for all $t \in [T]$.
With a matrix $A \in \mathbb{R}^{T\times T}$, this can be seen as $y_{t} = (Ax)_{t}=\sum_{k=0}^{T-1} a_{tk}x_{k}$ , where $(A)_{ij}=a_{ij} = h_{(i-j)\bmod T}$, for all $t \in [T]$.
Note that a matrix $A$ is a circular matrix.
From now on, we represent the filter with $A$ and the filtered signal with $y=Ax$.
We further assume that the $\ell_1$-norm of each row is normalized to $1$, i.e., $\|A_{t}\|_{1} = 1$ for all $t \in [T]$, where $A_{t}$ is $t$-th row of a matrix $A$.

As opposed to \textit{row-wise} subsampling in the past DP literature that can completely discard an individual's participation in a dataset with some probability, we contemplate \textit{column-wise} subsampling.
Formally, we define the subsampling operator as $\mathrm{Subsample}:\mathbb{R}^{T} \times \mathcal{J} \to \mathbb{R}^{T^\prime}$, 
where $\mathcal{J}=\{j \subseteq [T]\}$ is the power set of $[T]$ and $T^\prime$ is a length of subsampled signal.
We define $J \in \mathcal{J}$ as the random variable of subsampling indices.
We consider the Poisson sampling with parameter $p$ for $J$, i.e., for all $t\in[T]$, $t \in J$ with probability $p$ independently.
Therefore, we express the subsampling to $x$ with $j$ as $\hat{x} = \mathrm{Subsample}(x, j)$, where $\hat{x} \in \mathbb{R}^{|j|}$.

\section{Privacy Amplification by Subsampling}
Recalling the example of the California highway traffic data \citep{chen_freeway_2001}, the number of cars passed by the physical sensor is sampled once every five minutes.
The sensitivity grows exceedingly large with the number of sampled time steps, $T$; thus, the noise added to satisfy DP becomes too large to convey even daily or weekly trends for baseline mechanisms. 
However, it is extremely unlikely that a single individual participates at every sensor snapshot in a day.
Therefore, by making the mild assumption that an individual participates no more than at $I < T$ time steps in the signal, we can significantly reduce the sensitivity.
Without subsampling, this trivially reduces the ($L_2$-) sensitivity by a factor of $\sqrt{I/T}$. 
In this section, we demonstrate how we can reduce the sensitivity even further with high probability through subsampling and propose our $(\epsilon, \delta)$-DP algorithm that utilizes subsampling.

\subsection{Assumption}\label{sec:assumptions}
To make the above assumption precise, we formalize it as follows.\\
Denote $i$-th individual's participation to the dataset $D$ as $d_{i} \in \mathbb{R}^{T}$.
Thus, $x_t=\mathrm{count}(D)_t = \sum_{i} d_{i,t}$.
Without loss of generality, we assume $d_{i,t} \in \{0,1\}$ for all $i$ and $t$, i.e., an individual can participate at most once at each time step.
It is straightforward to extend our results to the case where participation is at most $M>1$ times.
The formal statement for limited individual's participation assumption is as follows.
\begin{assumption}\label{assumption:I}
    $\sup_i \sum_{t} d_{i,t} = I$, i.e., an individual can participate at most $I$ time steps.
\end{assumption}
    
\begin{table}[t]
    \centering
    \caption{Maximum number of check-ins per individual ($I$) for 15 most visited venues in Gowalla (left) and Foursquare (right) datasets.}
    \label{table:I}
    \begin{minipage}[t]{.45\columnwidth}
    \centering
    \begin{tabular}{lc}
        \hline
            ID &   $I$ ($T=725$)  \\
        \hline
          1 &  63 \\
          2 &  35 \\
          3 &  39 \\
          4 &  29 \\
          5 &  51 \\
          6 &  39 \\
          7 &  43 \\
          8 &  19 \\
          9 &  27 \\
         10 & 208 \\
         11 &  27 \\
         12 &  34 \\
         13 &  45 \\
         14 &  47 \\
         15 &  75 \\
        \hline
        \end{tabular}
    \end{minipage}
  \begin{minipage}[t]{.45\columnwidth}
    \centering
    \begin{tabular}{lc}
        \hline
        ID &   $I$ ($T=2126$) \\
        \hline
          1 & 208 \\
          2 & 127 \\
          3 &  98 \\
          4 & 225 \\
          5 & 265 \\
          6 & 121 \\
          7 & 191 \\
          8 &  78 \\
          9 &  94 \\
         10 & 182 \\
         11 & 170 \\
         12 & 206 \\
         13 &  94 \\
         14 &  82 \\
         15 & 350 \\
        \hline
    \end{tabular}
    \end{minipage} 
\end{table}
Assumption~\ref{assumption:I} formalizes the observation that no individuals participate at all time steps, especially when the signal is oversampled.
In most cases, the maximum time steps that an individual can participate at are smaller than $T$; thus, $I < T$.
This assumption is empirically evident in real time-series data.
Table~\ref{table:I} shows the maximum number of check-ins per individual for the top 15 most visited venues for Gowalla \citep{cho_friendship_2011} and Foursquare \citep{yang_nationtelescope_2015, yang_participatory_2016} datasets.
Similar to our highway example, it is implausible that an individual checks into a certain venue every time step.
We see that $I$ is, at least empirically, much smaller than $T$.
Since we do not know the exact value of the maximum individual's participation a priori, we must set $I$ conservatively in practice.
However, even if the assumption does not hold strictly, our guarantee is still valid with gentle privacy degradation as discussed in Section~\ref{sec:graceful}.
This assumption yields that $\Delta_{2,\mathrm{count}} = \sqrt{I}$.

Although the assumption reduces the $L_2$-sensitivity from $\sqrt{T}$ to $\sqrt{I}$, we will show that random subsampling reduces it further in the following sections.

\subsection{Sensitivity Reduction by Subsampling}\label{sec:sens_analysis}
With the assumption in mind, the naive approach to achieve DP is to apply the Gaussian mechanism with the sensitivity parameter $\Delta_{2,\mathrm{count}} = \sqrt{I}$.
The dependency on $\sqrt{I}$ is undesirable as $I$, which tends to be proportional to $T$, becomes larger. 
Therefore, we propose to apply random subsampling to relax this dependency.
The intuition behind it is that the probability that subsampled time steps contain all (or most) of an individual's participation is very low under Assumption~\ref{assumption:I}; thus, we get the further dependency reduction.
Formally, we show that $\mathrm{Subsample}$ helps reduce the $L_2$-sensitivity with high probability, where the probability is over the choice of $J$, under Assumption~\ref{assumption:I}.

Throughout the analysis, we assume that $D = D^\prime \cup \{d_i\}$, i.e., $i$-th individual is the only one who participates in $D$ and does not in $D^\prime$.
Also, we let $j_i = \{t | d_{i,t} =1\}$, time indices that $i$-th individual participates.

\subsubsection{Sensitivity Reduction by Subsampling Only}
We start from examining the sensitivity of $f_{\mathrm{s}}(\cdot,\cdot) = \mathrm{Subsample}(\mathrm{count}(\cdot), \cdot)$.
Intuitively, under Assumption~\ref{assumption:I}, the probability that $J$ contains almost all $I$ time steps is small; thus, we obtain the sensitivity reduction with high probability.
\begin{theorem}\label{th:ss}
Fix $I^\prime \leq I$. 
Let $\delta^\prime=\Pr\{|J \cap j_i| > I^\prime\} = \sum_{t =I^\prime+1}^{I} {I \choose t} p^{t} (1-p)^{I-t}$.
Then, with probability $1-\delta^\prime$, $\Delta_{2,f_\mathrm{s}}=\sqrt{I^\prime}$.
\end{theorem}
The proof follows directly from the tail probability of the binomial distribution with parameters $I$ and $p$.
By using the tail bound of the binomial distribution, we can explicitly express the upper bound of $I^\prime$ for fixed $\delta^\prime$ with $I$ and $p$.
\begin{corollary}\label{co:tail_Ip}
Let $\delta^\prime < 1$.
Then, with probability $1-\delta^\prime$, $\Delta_{2,f_\mathrm{s}}=\sqrt{I^\prime}\leq \sqrt{I p + \sqrt{\frac{I}{2} \log 1/\delta^\prime}}$.
\end{corollary}
The claim is a direct consequence of the Hoeffding's inequality.
We see with high probability that the dependency on $I$ is replaced by $Ip$ plus the probably small additional term.

\subsubsection{Sensitivity Reduction by Filtering and Subsampling}
Next, we consider the case where we filter the signal before subsampling.
Some filters have an averaging effect across time steps; thus, filtered signals become smoother and less noisy in general.

Again, we go back to our running example of the CA highway traffic data.
The signal is very oversampled for an analyst who is only interested in how traffic trends change every two hours.
Changes on the order of five minutes do not contribute to their utility, but magnifies the sensitivity significantly.
By filtering, we effectively average in time to smooth the signal, and ultimately reduce the variance after subsampling.
As detailed in Section~\ref{sec:prelim} we represent a filter as a matrix $A$, which, after multiplying with the signal, produces the smoother signal.

However, from the DP point of view, filtering spreads the content of $t$-th time step across all time steps; thus, the effects of subsampling are now less obvious.
As an extreme example, consider a filter that simply averages across all time points, i.e., $A_{ij}=1/T$ for any $i$ and $j$.
Then, any individual's contribution scatters equally along all time steps.
Thus, the intuition above does not hold here, but still, we can apply a similar way of thinking if $A$ is somewhat similar to the identity matrix---there exist time steps that $i$-th individual's influence is large, and the probability of choosing almost all such time steps is low.

We define the function as $f_{\mathrm{fs}}(\cdot,\cdot) = \mathrm{Subsample}(A\;\mathrm{count}(\cdot), \cdot)$.
By letting $\delta_i = 1$ if $i\in J$ and $0$ otherwise and
$B=\mathrm{diag}(\delta_1,\ldots, \delta_T)A$, we can express $f_{\mathrm{fs}}(x, J) = Bx$.
Fix two signals $x,x^\prime$ induced by neighboring datasets.
Then, it holds that
\begin{align*}
    \Delta_{2, f_{\mathrm{fs}}}^{2} &= \|Bx-Bx^\prime\|_2^2\\
    &= (x-x^\prime)^\top B^\top B (x-x^\prime)\\
    &\leq \lambda_{\max} (B^\top B) \|x-x^\prime\|^2_2
    \leq \sigma_{\max}^2 (B) I,
\end{align*}
where the last inequality follows from Assumption~\ref{assumption:I}.
It remains to bound the maximum singular value of $B$ to complete bounding the sensitivity.\\
To bound how large the maximum singular value is likely to be, we use the following simplified version of the theorem by \citet{tropp_introduction_2015}.
\begin{theorem}[Simplified version, \citep{tropp_introduction_2015}]
For $\varepsilon \geq 0$,
\begin{align*}
    \mathbb{P}\left\{\sigma^{2}_{\max }(B) \geq(1+\varepsilon) p \sigma^{2}_{\max}(A)\right\} 
    \leq 2\mathrm{srank}(A)\left(\frac{\mathrm{e}^{\varepsilon}}{(1+\varepsilon)^{1+\varepsilon}}\right)^{p\sigma^{2}_{\max }(A) / L},
\end{align*}
where $\mathrm{srank}(A) = \frac{\|A\|_{F}^{2}}{\|A\|^{2}}$ is the stable rank of $A$, the ratio between the squared Frobenius norm of $A$ and the squared spectral norm of $A$,
and $L=\max_t \|A_{t}\|_2^2$.
\end{theorem}

As we assume in Section~\ref{sec:prelim} that $A$ is a circular matrix and $\|A_t\|_{1} = 1$ for all $t\in [T]$, $\sigma_{\max }(A) = 1$ and $L=\|A_{1}\|_2^2$.
Combining this theorem with the sensitivity bound with the maximum singular value, we are able to bound the reduction in sensitivity from \emph{both} subsampling \emph{and} filtering simultaneously.
Our final sensitivity reduction theorem for filtered and subsampled signals is given below.
\begin{theorem}\label{th:fss}
Fix $\alpha \in (0,1]$. 
Let $\delta^\prime = 2\mathrm{srank}(A)\left(\frac{e^{\alpha^2/p-1}}{(\alpha^2/p)^{\alpha^2/p}}\right)^{p/L}$.
Then, with probability at least $1-\delta^\prime$,
$\Delta_{2, f_{\mathrm{fs}}} = \alpha \sqrt{I}$.
\end{theorem}

Note that the theorem above does not fully make use of Assumption~\ref{assumption:I}; thus, the statement holds even if we replace $I$ with $T$.
Also, we note that by taking $A$ to be an identity matrix, the function $f_\mathrm{fs}$ is exactly the same as $f_\mathrm{s}$, but the bound is looser than Theorem~\ref{th:ss} in general.
Furthermore, it is hard to obtain the explicit upper bound for $\alpha$ in general, but we give the relationship between $\alpha$ and $\delta^\prime$ in Appendix~3.

\subsection{Algorithm}\label{sec:alg}
\begin{algorithm}[tb]
  \caption{Our $(\epsilon, \delta)$-DP Algorithm}
  \label{alg:fss}
\begin{algorithmic}
  \STATE {\bfseries Input:} count signal $x\in\mathbb{R}^T$, filter $A\in\mathbb{R}^{T\times T}$, parameters $p$, $\alpha$, $\epsilon$, $\delta$ 
  \STATE {\bfseries Output:} $\hat{z}\in\mathbb{R}^T$
  \STATE $y=Ax$
  \STATE Sample $J$ uniform randomly with a parameter $p$.
  \STATE $y_{\mathrm{s}} = \mathrm{Subsample}(y, J)$
  \STATE $z = y_{\mathrm{s}} + (\xi_1,\ldots,\xi_{|J|})^\top$, where $\xi_i\sim\mathcal{N}\left(0, \frac{2\ln (1.25/\delta)}{(\epsilon / (\alpha \sqrt{I}))^{2}}\right)$
  \STATE Interpolate $z$ to get $\hat{z}$ with $J$. 
\end{algorithmic}
\end{algorithm}

Algorithm~\ref{alg:fss} gives our $(\epsilon, \delta)$-DP algorithm for a filtered and subsampled count signal. 
The central idea is to utilize the sensitivity analysis in the previous section and add less variance noise than the standard Gaussian mechanism.
One note is that the interpolation in the last line of Algorithm~\ref{alg:fss} does not incur a privacy breach by using $J$ as long as the interpolation is bijective given $J$, which is usually the case.

From Theorems~\ref{th:ss} and~\ref{th:fss}, we see that Algorithm~\ref{alg:fss} satisfies $(\epsilon,\delta)$-DP.
In short, our privacy guarantee allows smaller $\epsilon$ in exchange for a little larger $\delta$ under the same noise variance as the standard Gaussian mechanism.
\begin{corollary} \label{col:fss}
Let $\delta^\prime$ be the one in Theorem~\ref{th:fss}.
Algorithm~\ref{alg:fss} satisfies 
$(\epsilon, \delta+\delta^{\prime}(\exp(\epsilon/\alpha)-\exp(\epsilon)))$-DP.
\end{corollary}
The corollary follows directly from Theorem~\ref{th:fss}.
We provide the detailed proof in Appendix~1.\\
The guarantee implicitly requires $\delta^\prime$ to be small enough.
In other words, if $\delta^\prime(\exp(\epsilon/\alpha)-\exp(\epsilon))$ is smaller or equal to $\delta$, the compensation for the $\delta$ term is negligible and we obtain the better privacy.
Also, $\delta^\prime$ is a monotonously decreasing function with respect to $\alpha$; thus, one can easily find appropriate $\alpha$ to satisfy desired $\delta^\prime$ numerically.

When $A$ is an identity matrix, i.e., not filtering before subsampling, we have a better privacy guarantee.
\begin{corollary}
Let $A$ be an identity matrix, $I^\prime = \left\lceil \alpha^2 I \right\rceil$
and $\delta^\prime$ be the one in Theorem~\ref{th:ss}.
Algorithm~\ref{alg:fss} satisfies 
$(\epsilon, \delta+\delta^{\prime}(\exp(\sqrt{I/I^\prime}\epsilon)-\exp(\epsilon)))$-DP.
\end{corollary}
The proof is almost identical to the one for Corollary~\ref{col:fss}.
Note that the above guarantee heavily relies on Assumption~\ref{assumption:I} as opposed to Corollary~\ref{col:fss}.

\subsubsection{Graceful Privacy Degradation When Assumption~\ref{assumption:I} Fails}\label{sec:graceful}
Although we succeed in amplifying privacy by subsampling, one would cast doubt on the validity of Assumption~\ref{assumption:I}, i.e., any individual participates at most $I$ time steps.
Here, we show even if the assumption does not hold precisely, our algorithm still satisfies DP and the privacy level does not degrade abruptly.
\begin{proposition}
    If $\sup_i \sum_t d_{i,t} = cI$ for some $c>1$,
    then Algorithm~\ref{alg:fss} satisfies $(\sqrt{c}\epsilon, \delta+\delta^\prime (\exp(\sqrt{c}\epsilon/\alpha) - \exp(\sqrt{c}\epsilon))$-DP.
\end{proposition}
In short, the above proposition claims that if the \emph{actual} participation limit is $cI$ instead of $I$, privacy parameters increase only slightly; in particular, $\epsilon$ only increases by a factor of $\sqrt{c}$.

\section{Experiments}
We next investigate how well our algorithm works in practice by empirically evaluating it on real and synthetic data. 
Specifically, we ask the following questions:
\begin{enumerate}
    \item How does our algorithm compare with existing baselines such as the Gaussian mechanism and DFT in terms of accuracy?
    \item How does filtering before subsampling improve accuracy?
    \item How does the sampling frequency of the time-series data affect accuracy?
\end{enumerate}
We answer the first question by experimenting with three real datasets in Section~\ref{sec:real_exp}.
We then answer the rest of the questions with a synthetic dataset in Section~\ref{sec:synth_exp}.

\subsection{Setup}

\paragraph{Datasets.} We consider four datasets -- three real, and one synthetic. 

\textbf{PeMS} ($T=1800$) \citep{chen_freeway_2001} is a dataset collected by California Transportation Agencies (CalTrans) Performance Measurement System (PeMS).
We select a sensor from a highway in the Bay area, California, and sample the flow data, the number of cars that passed the sensor, every five minutes from Jan. 2017.
\textbf{Gowalla} ($T=725$) \citep{cho_friendship_2011} is a dataset of check-in information from a location-based social networking website called Gowalla. We use the most visited location and aggregate the check-in count every 24 hours from Feb. 2009.
\textbf{Foursquare} ($T=2126$) \citep{yang_nationtelescope_2015, yang_participatory_2016} is a similar dataset with check-in information from the location data platform Foursquare. We pick the most checked-in place and aggregate the count every 6 hours from April 2012. Since raw Gowalla and Foursquare datasets contain user IDs, we use them to calculate the empirical $I$ in Table~\ref{table:I}. We set $I = \left\lceil T/10 \right\rceil$ as confirmed by Table~\ref{table:I}. 

Since we do not have the data generating mechanisms for real data, to understand how the sampling frequency affects the accuracy and the conditions under which filtering helps, we also experiment with a synthetic dataset \textbf{Synth}. 
We generate the signal value given time step $t$ as $x_t = a\sin(\omega t) + b + ct$, where we set $a=200$, $b=500$ and $c=0.1$.
The resulting signal has two overall trends: periodicity and linear growth.
We expect a good algorithm to return a randomized version that maintains \emph{both} trends.
We control the sampling frequency to examine how the extent of oversampling affects the utility.
We do this by, given (relative) sampling frequency $f$, generating the signal $x^{(f)} = (x_{0}, x_{1/f}, x_{2/f}, \ldots)$.
Thus, as the relative frequency increases, the signal is oversampled more and the length gets larger.
We set $T=10000$ for $f=1$ (thus, $T=10000\cdot f$ given $f$) and $I=100$ for any $f$.
We generate several signals with varying $f$'s ($f\in \{1, 1/2, 1/4, 1/8, 1/16, 1/32, 1/64\}$). 
We provide the reason why we fix $I$ for the Synth data in Appendix~2.
Furthermore, we consider the case where there is observation noise: unbiased randomness added to each time step that is typical in real-world time-series. To do so, we add mean-zero Gaussian noise to each time step, yielding a more realistic, noisy version of the signal: $\tilde{x}_t = x_t + d\hat{\sigma}_t$, where $\hat{\sigma}_t$'s are drawn i.i.d. from $\mathcal{N}(0,1)$ and $d=100$.

\paragraph{Algorithms.} 
The simplest baseline for our problem is the standard Gaussian mechanism.
The Gaussian mechanism for time-series data adds noise corresponding to the sensitivity $\sqrt{I}$ to each of $T$ time steps by considering time-series data as a $T$-dimensional vector.
The more involved baseline is the algorithm by \citet{rastogi_differentially_2010} (DFT).
It adds noise to the first $k$ discrete Fourier transform coefficients and ignores higher frequency contents of the signal.
For DFT, we use the hyperparameters reported as being the best in their paper. While DFT is originally an $\epsilon$-DP algorithm, we extend it to be $(\epsilon, \delta)$-DP by changing the noise distribution to Gaussian from Laplace for direct comparison with our algorithm. 
For our algorithm, we consider two variations---with and without filtering.

\paragraph{Experiment Setup.} We set the privacy parameters and $\alpha$ so that each mechanism guarantees $(\epsilon,\delta)=(0.5, 10^{-4})$-DP throughout the experiments.
Accuracy is measured by the mean absolute error (MAE) between the raw signal and the output.
For the noisy Synth data, we measure the MAE between the true signal (before adding observation noise) and the output.
For each time-series, we repeat the algorithms $1000$ times and report the mean and standard deviation of MAE; this is because the outputs depend heavily on the randomness of the mechanisms.
We choose a subsampling parameter for our algorithm $p$ to be $0.1$, i.e., the subsampled signal length would be around $T/10$ for the real datasets.
For the synthetic signal, $p=0.1 \cdot (1-\log(f)/4)$.
We present further details of the experiments in Appendix~2.

\subsection{Results on Real Data} \label{sec:real_exp}
\begin{table*}[t]
\caption{Mean MAEs and standard deviations of Gaussian mechanism, DFT and our algorithms (with and without filter) on PeMS, Gowalla, and Foursquare dataset for $(\epsilon,\delta)=(0.5,10^{-4})$.
We see that our algorithm without filtering performs the best on these real world datasets. We explore when filtering is optimal on the Synth dataset.}
\label{table:res}
\begin{center}
\begin{tabular}{l|ccc}
\hline
 & PeMS & Gowalla & Foursquare \\
\hline
Gaussian  & 93.0 $\pm$ 1.6   & 33.0 $\pm$ 1.5 & 57.4 $\pm$ 1.5 \\
DFT  & 55.7 $\pm$ 2.8  & 39.3 $\pm$ 3.5 & 42.9 $\pm$ 3.4\\
\textbf{Ours w/o Filter} & 42.8 $\pm$ 2.9  & 16.7 $\pm$ 2.7  & 29.5 $\pm$ 2.4 \\
\textbf{Ours w/ Filter}  & 60.9 $\pm$ 4.0  & 21.2 $\pm$ 3.9 & 39.6 $\pm$ 3.6 \\
\hline
\end{tabular}
\end{center}
\end{table*}

Table~\ref{table:res} summarizes the MAEs on the real datasets.
We see that for all the datasets, our algorithm without filtering performs the best, indicating the superiority of this algorithm. 
Ours with filtering performs better than the baselines on the Gowalla and Foursquare dataset, but slightly worse than DFT on the PeMS dataset.
This might be due to the periodic nature of the PeMS signal---DFT is suitable for periodic signals in general.
Although the overall performance of our algorithms is better than the baselines, it seems that the variance of MAE is slightly larger than the standard Gaussian mechanism. 
We anticipate that the main reason is the interpolation in Algorithm~\ref{alg:fss}.

\subsection{Results on Synthetic Data}\label{sec:synth_exp}
\begin{figure*}[t]
\centering
\includegraphics[width=\textwidth]{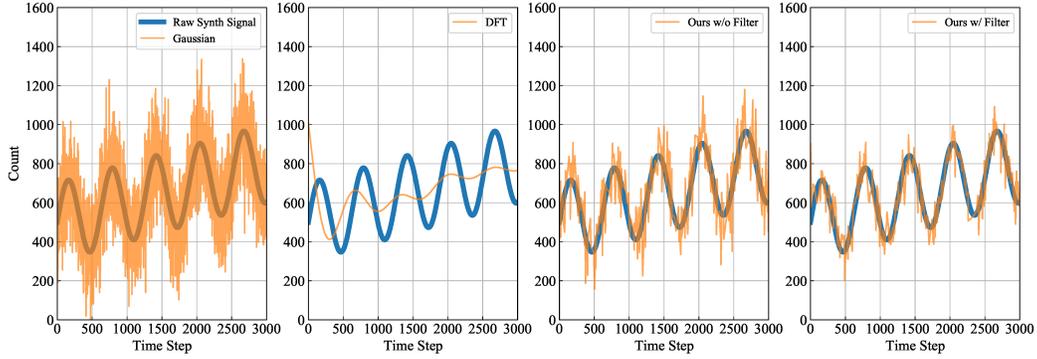}
\caption{
Two baseline mechanisms (left) and our mechanism with and without filtering (right), each operating on Synth time-series with observation noise. We plot the noiseless Synth time-series in blue for reference. 
Output by our algorithm with filtering concentrates around raw signal the most, demonstrating value of our filter $+$ subsample algorithm when operating on noisy time-series. 
}
\label{fig:synth_signal_noise}
\end{figure*}

\begin{figure*}[t]
\centering
\begin{subfigure}[b]{0.42\textwidth}
  \centering
  \includegraphics[width=\textwidth]{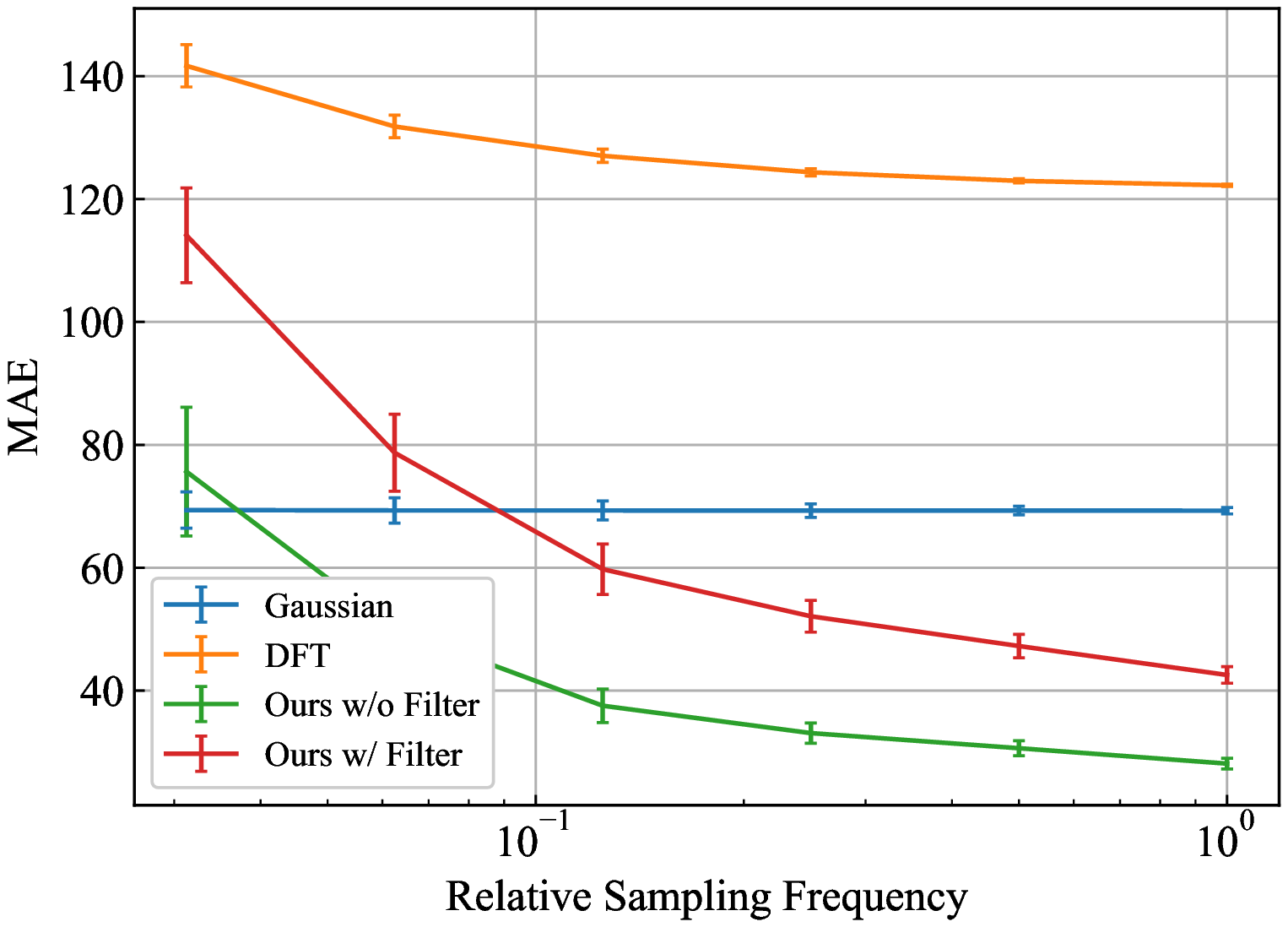}
  \caption{Without observation noise}
  \label{fig:synth_noiseless}
\end{subfigure}
\hfill
\begin{subfigure}[b]{0.42\textwidth}
  \centering
  \includegraphics[width=\textwidth]{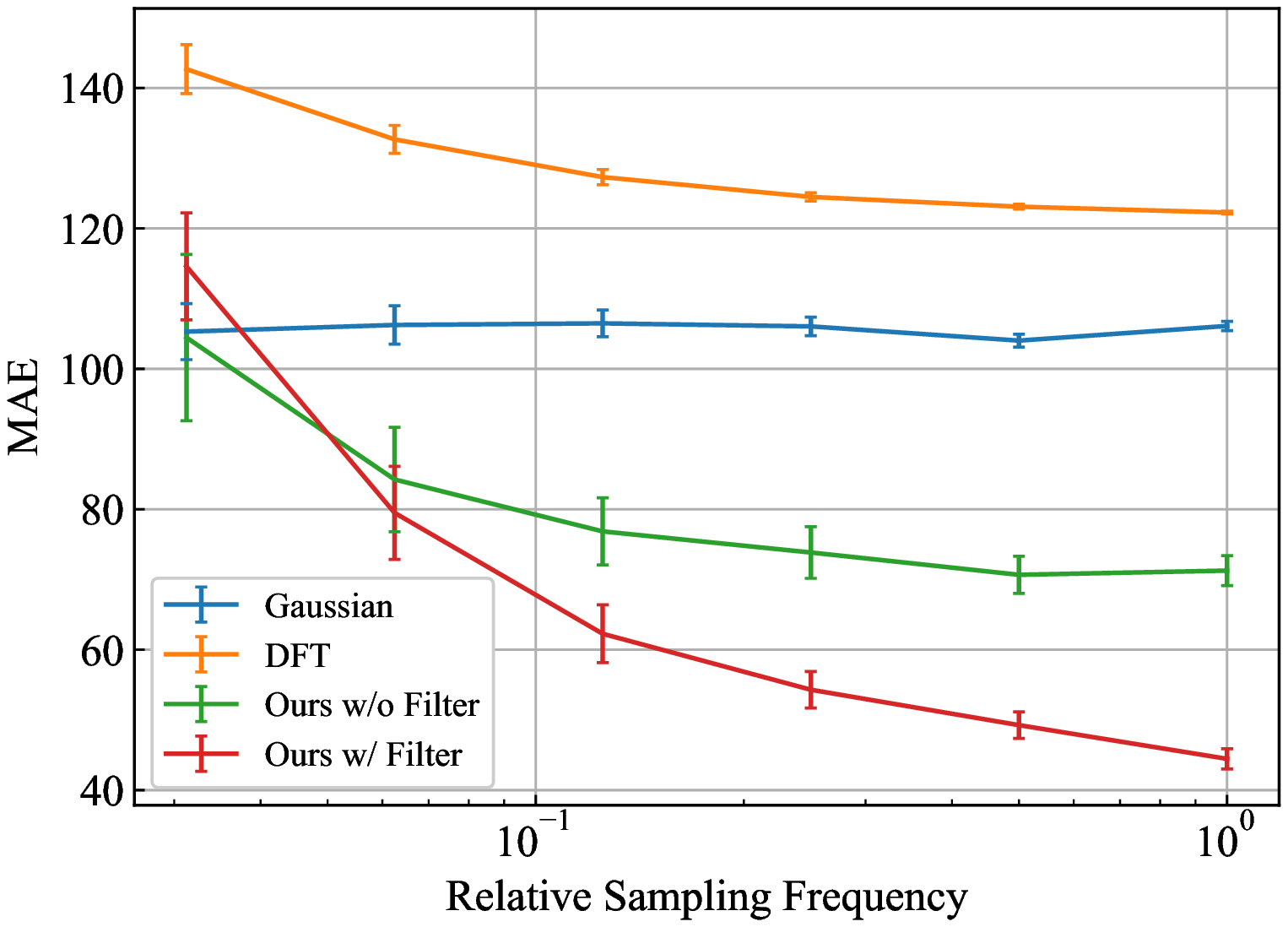}
  \caption{With observation noise}
  \label{fig:synth_noise}
\end{subfigure}
\caption{
Mean MAEs (smaller is better) and standard deviations of Gaussian mechanism, DFT, and our algorithms (with and without filter) on Synth data under varying relative sampling frequencies for $(\epsilon,\delta)=(0.5,10^{-4})$.
Signal is sampled more frequently, i.e., more oversampled, as relative sampling frequency gets larger.
We see that \textbf{(1)} our algorithms (with and without filter) perform better as signal gets more oversampled, and \textbf{(2)} our algorithm with filtering performs best when the input signal is noisy (right). 
}
\end{figure*}

To better understand when filtering works well, we compare the results with and without observation noise. We anticipate that the averaging effect of filtering offers better performance when observation noise is present. 
Figure~\ref{fig:synth_signal_noise} shows in orange examples of outputs of four mechanisms: the two baseline mechanisms and our mechanism with and without filtering, each operating on the Synth time-series with observation noise. For reference the Synth signal is plotted in blue (without obs. noise for clarity). 
We show the outputs operating on the \emph{noiseless} Synth signal in Appendix~2.
Figures~\ref{fig:synth_noiseless} and~\ref{fig:synth_noise} show the MAEs for Synth dataset with and without observation noise while sweeping relative sampling frequency $f$. 
We see that our mechanism with filtering generally outperforms the one without filtering when there is observation noise, confirming our hypothesis.
Filtering before subsampling has an averaging effect in time, canceling out the observation noise and thus accurately conveying the underlying trends.
We can observe this effect directly in Figure~\ref{fig:synth_signal_noise}. The output from our algorithm with filtering concentrates around the raw Synth signal the most, yielding a lower MAE and better capturing underlying trends.
On the other hand, when the signal is noiseless, filtering does not help achieve better utility.
As seen in Section~\ref{sec:sens_analysis}, the sensitivity reduction is larger for subsampling without filtering; thus, ours without filtering is more helpful for less noisy signals.

To understand the impact of oversampling, i.e., how sampling frequency matters to utility, we observe the results under several (relative) sampling frequencies ($f$).
The larger $f$ is, the more oversampled the signal is.
For both Figures~\ref{fig:synth_noiseless} and~\ref{fig:synth_noise}, we see a similar trend that MAEs get smaller as $f$ becomes larger for our algorithms (with and without filtering).
This trend validates the hypothesis that almost no information gets lost by subsampling for oversampled signals.
In such a case, we gain utility due to the reduced sensitivity achieved by subsampling.
By contrast, for less frequently sampled, fast-moving signals, our algorithms perform worse than the standard Gaussian mechanism.
We expect that this is because the error caused by the interpolation dominates the error by adding noise.
The linear interpolation does not supplement the information lost by subsampling.
Note that MAEs for the Gaussian mechanism and DFT are almost constant because, for experiments on Synth, we fix $I$ for all $f$'s.

\subsection{Discussion}
Throughout the experiments, we have made the three questions clear.

Our experimental results on real data reveal that our algorithm, especially without filtering, is superior to the baseline mechanisms, the Gaussian mechanism and DFT, in terms of MAE.
This is mainly because the sensitivity reduction by subsampling reduces the noise variance and the error incurred by the interpolation is small due to the oversampled nature of real data.

Filtering improves utility when signals contain observation noise as confirmed by the result on the Synth data.
It is helpful when we want to capture underlying trends rather than to reproduce original signals because applying a filter has an averaging effect and thus lessens the effect of observation noise.
On the contrary, for noiseless signals, we should use our algorithm without filtering since the sensitivity reduction is larger for subsampling but not filtering in general.

The result on the Synth data also indicates that our algorithm works better as data gets oversampled.
This result verifies our hypothesis that subsampling preserves overall trends for oversampled signals.
The sensitivity reduction is also significant since our algorithm outperforms the baseline mechanisms for such signals.

\section{Conclusion and Future Work}
This paper investigates how subsampling and filtering help achieve a better privacy-utility tradeoff for the longstanding problem of publishing sanitized aggregate time-series data with a DP guarantee. 
Using a novel concentration analysis, we show subsampling and filtering reduce the sensitivity with high probability under reasonable assumptions for real time-series data. 
We then propose a class of DP mechanisms exploiting subsampling and/or filtering, which empirically outperform baselines. 
Going forward, we plan to develop theoretical utility guarantees for our mechanisms, and explore how alternative interpolation schemes impact utility.
Finally, we believe feature-wise subsampling is beneficial beyond time-series data, and are investigating other applications for our techniques.

\section*{Acknowledgments}
TK, CM, and KC would like to thank ONR under N00014-20-1-2334 and UC Lab Fees under LFR 18-548554 for research support. 
Also, TK is supported in part by Funai Overseas Fellowship.
We would also like to thank our reviewers for their insightful feedback.

\newpage
\bibliography{ms}

\begin{thebibliography}{}

\bibitem[Abadi et~al., 2016]{abadi_deep_2016}
Abadi, M., Chu, A., Goodfellow, I., McMahan, H.~B., Mironov, I., Talwar, K.,
  and Zhang, L. (2016).
\newblock Deep {{Learning}} with {{Differential Privacy}}.
\newblock In {\em Proceedings of the 2016 {{ACM SIGSAC Conference}} on
  {{Computer}} and {{Communications Security}}}, {{CCS}} '16, pages 308--318,
  {New York, NY, USA}. {Association for Computing Machinery}.

\bibitem[Balle et~al., 2018]{balle_privacy_2018}
Balle, B., Barthe, G., and Gaboardi, M. (2018).
\newblock Privacy {{Amplification}} by {{Subsampling}}: Tight {{Analyses}} via
  {{Couplings}} and {{Divergences}}.
\newblock In {\em Advances in {{Neural Information Processing Systems}}},
  volume~31. {Curran Associates, Inc.}

\bibitem[Bassily et~al., 2014]{bassily_private_2014}
Bassily, R., Smith, A., and Thakurta, A. (2014).
\newblock Private {{Empirical Risk Minimization}}: Efficient {{Algorithms}} and
  {{Tight Error Bounds}}.
\newblock In {\em 2014 {{IEEE}} 55th {{Annual Symposium}} on {{Foundations}} of
  {{Computer Science}}}, pages 464--473.

\bibitem[Beimel et~al., 2014]{beimel_bounds_2014}
Beimel, A., Brenner, H., Kasiviswanathan, S.~P., and Nissim, K. (2014).
\newblock Bounds on the sample complexity for private learning and private data
  release.
\newblock {\em Mach Learn}, 94(3):401--437.

\bibitem[Beimel et~al., 2013]{beimel_characterizing_2013}
Beimel, A., Nissim, K., and Stemmer, U. (2013).
\newblock Characterizing the sample complexity of private learners.
\newblock In {\em Proceedings of the 4th Conference on {{Innovations}} in
  {{Theoretical Computer Science}}}, {{ITCS}} '13, pages 97--110, {New York,
  NY, USA}. {Association for Computing Machinery}.

\bibitem[Cao et~al., 2019]{cao_priste_2019}
Cao, Y., Xiao, Y., Xiong, L., and Bai, L. (2019).
\newblock {{PriSTE}}: From {{Location Privacy}} to {{Spatiotemporal Event
  Privacy}}.
\newblock In {\em 2019 {{IEEE}} 35th {{International Conference}} on {{Data
  Engineering}} ({{ICDE}})}, pages 1606--1609, {Macao, Macao}. {IEEE}.

\bibitem[Cao et~al., 2017]{cao_quantifying_2017}
Cao, Y., Yoshikawa, M., Xiao, Y., and Xiong, L. (2017).
\newblock Quantifying {{Differential Privacy}} under {{Temporal Correlations}}.
\newblock In {\em 2017 {{IEEE}} 33rd {{International Conference}} on {{Data
  Engineering}} ({{ICDE}})}, pages 821--832.

\bibitem[Chan et~al., 2011]{chan_private_2011}
Chan, T.-H.~H., Shi, E., and Song, D. (2011).
\newblock Private and {{Continual Release}} of {{Statistics}}.
\newblock {\em ACM Trans. Inf. Syst. Secur.}, 14(3):26:1--26:24.

\bibitem[Chen et~al., 2001]{chen_freeway_2001}
Chen, C., Petty, K., Skabardonis, A., Varaiya, P., and Jia, Z. (2001).
\newblock Freeway {{Performance Measurement System}}: Mining {{Loop Detector
  Data}}.
\newblock {\em Transportation Research Record}, 1748(1):96--102.

\bibitem[Cho et~al., 2011]{cho_friendship_2011}
Cho, E., Myers, S.~A., and Leskovec, J. (2011).
\newblock Friendship and mobility: User movement in location-based social
  networks.
\newblock In {\em Proceedings of the 17th {{ACM SIGKDD}} International
  Conference on {{Knowledge}} Discovery and Data Mining}, {{KDD}} '11, pages
  1082--1090, {New York, NY, USA}. {Association for Computing Machinery}.

\bibitem[Dwork et~al., 2006a]{dwork_our_2006}
Dwork, C., Kenthapadi, K., McSherry, F., Mironov, I., and Naor, M. (2006a).
\newblock Our {{Data}}, {{Ourselves}}: Privacy {{Via Distributed Noise
  Generation}}.
\newblock In Vaudenay, S., editor, {\em Advances in {{Cryptology}} -
  {{EUROCRYPT}} 2006}, Lecture {{Notes}} in {{Computer Science}}, pages
  486--503, {Berlin, Heidelberg}. {Springer}.

\bibitem[Dwork et~al., 2006b]{dwork_calibrating_2006}
Dwork, C., McSherry, F., Nissim, K., and Smith, A. (2006b).
\newblock Calibrating {{Noise}} to {{Sensitivity}} in {{Private Data
  Analysis}}.
\newblock In Halevi, S. and Rabin, T., editors, {\em Theory of
  {{Cryptography}}}, Lecture {{Notes}} in {{Computer Science}}, pages 265--284,
  {Berlin, Heidelberg}. {Springer}.

\bibitem[Dwork et~al., 2010]{dwork_differential_2010}
Dwork, C., Naor, M., Pitassi, T., and Rothblum, G.~N. (2010).
\newblock Differential privacy under continual observation.
\newblock In {\em Proceedings of the Forty-Second {{ACM}} Symposium on
  {{Theory}} of Computing}, {{STOC}} '10, pages 715--724, {New York, NY, USA}.
  {Association for Computing Machinery}.

\bibitem[Dwork and Roth, 2014]{dwork_algorithmic_2014}
Dwork, C. and Roth, A. (2014).
\newblock The {{Algorithmic Foundations}} of {{Differential Privacy}}.
\newblock {\em Found. Trends Theor. Comput. Sci.}, 9(3\textendash 4):211--407.

\bibitem[Fan and Xiong, 2012]{fan_real-time_2012}
Fan, L. and Xiong, L. (2012).
\newblock Real-time aggregate monitoring with differential privacy.
\newblock In {\em Proceedings of the 21st {{ACM}} International Conference on
  {{Information}} and Knowledge Management}, {{CIKM}} '12, pages 2169--2173,
  {New York, NY, USA}. {Association for Computing Machinery}.

\bibitem[Kasiviswanathan et~al., 2011]{kasiviswanathan_what_2011}
Kasiviswanathan, S.~P., Lee, H.~K., Nissim, K., Raskhodnikova, S., and Smith,
  A. (2011).
\newblock What {{Can We Learn Privately}}?
\newblock {\em SIAM J. Comput.}, 40(3):793--826.

\bibitem[Meehan and Chaudhuri, 2021]{meehan_location_2021}
Meehan, C. and Chaudhuri, K. (2021).
\newblock Location {{Trace Privacy Under Conditional Priors}}.
\newblock In {\em Proceedings of {{The}} 24th {{International Conference}} on
  {{Artificial Intelligence}} and {{Statistics}}}, pages 2881--2889. {PMLR}.

\bibitem[Ny and Pappas, 2014]{ny_differentially_2014}
Ny, J.~L. and Pappas, G.~J. (2014).
\newblock Differentially {{Private Filtering}}.
\newblock {\em IEEE Transactions on Automatic Control}, 59(2):341--354.

\bibitem[Rastogi and Nath, 2010]{rastogi_differentially_2010}
Rastogi, V. and Nath, S. (2010).
\newblock Differentially private aggregation of distributed time-series with
  transformation and encryption.
\newblock In {\em Proceedings of the 2010 International Conference on
  {{Management}} of Data - {{SIGMOD}} '10}, page 735, {Indianapolis, Indiana,
  USA}. {ACM Press}.

\bibitem[Tropp, 2015]{tropp_introduction_2015}
Tropp, J.~A. (2015).
\newblock An {{Introduction}} to {{Matrix Concentration Inequalities}}.
\newblock {\em MAL}, 8(1-2):1--230.

\bibitem[Xiao and Xiong, 2015]{xiao_protecting_2015}
Xiao, Y. and Xiong, L. (2015).
\newblock Protecting {{Locations}} with {{Differential Privacy}} under
  {{Temporal Correlations}}.
\newblock In {\em Proceedings of the 22nd {{ACM SIGSAC Conference}} on
  {{Computer}} and {{Communications Security}}}, {{CCS}} '15, pages 1298--1309,
  {New York, NY, USA}. {Association for Computing Machinery}.

\bibitem[Yang et~al., 2015]{yang_nationtelescope_2015}
Yang, D., Zhang, D., Chen, L., and Qu, B. (2015).
\newblock {{NationTelescope}}: Monitoring and visualizing large-scale
  collective behavior in {{LBSNs}}.
\newblock {\em Journal of Network and Computer Applications}, 55:170--180.

\bibitem[Yang et~al., 2016]{yang_participatory_2016}
Yang, D., Zhang, D., and Qu, B. (2016).
\newblock Participatory {{Cultural Mapping Based}} on {{Collective Behavior
  Data}} in {{Location}}-{{Based Social Networks}}.
\newblock {\em ACM Trans. Intell. Syst. Technol.}, 7(3):30:1--30:23.

\end{thebibliography}


\begin{thebibliography}{}

\bibitem[Virtanen et~al., 2020]{virtanen_scipy_2020}
Virtanen, P., Gommers, R., Oliphant, T.~E., Haberland, M., Reddy, T.,
  Cournapeau, D., Burovski, E., Peterson, P., Weckesser, W., Bright, J., {van
  der Walt}, S.~J., Brett, M., Wilson, J., Millman, K.~J., Mayorov, N., Nelson,
  A. R.~J., Jones, E., Kern, R., Larson, E., Carey, C.~J., Polat, {\.I}., Feng,
  Y., Moore, E.~W., VanderPlas, J., Laxalde, D., Perktold, J., Cimrman, R.,
  Henriksen, I., Quintero, E.~A., Harris, C.~R., Archibald, A.~M., Ribeiro,
  A.~H., Pedregosa, F., and {van Mulbregt}, P. (2020).
\newblock {{SciPy}} 1.0: Fundamental algorithms for scientific computing in
  {{Python}}.
\newblock {\em Nature Methods}, 17(3):261--272.

\end{thebibliography}

\end{document}

\section{Proof of Results}
\subsection{Proof of Theorem 2}
\begin{proof}
For fixed $I^\prime \leq I$ ($I^\prime \in \mathbb{N}$), it holds that 
$\Pr \{\Delta_{2,f_\mathrm{s}} > \sqrt{I^\prime}\} = \Pr\{|J \cap j_i| > I^\prime\}$,
where the probability is taken over the randomness of $J$.
Since $|J \cap j_i|$ follows the Binomial distribution with parameters $|j_i|$ and $p$, the tail probability is bounded as below:
\begin{align*}
    \Pr\{|J \cap j_i| > I^\prime\} &= \sum_{t =I^\prime+1}^{|j_i|} {|j_i| \choose t} p^{t} (1-p)^{|j_i|-t}\\
    &\leq \sum_{t =I^\prime+1}^{I} {I \choose t} p^{t} (1-p)^{I-t},
\end{align*}
where the last inequality follows from $|j_i| \leq I$ and the monotonicity in a parameter $n$ of the tail probability of the Binomial distribution.
Setting the last term to be $\delta^\prime$ concludes the proof.
\end{proof}

\subsection{Proof of Corollary 1}
\begin{proof}
From Theorem~2, for fixed $\delta^\prime$, if $I^\prime$ satisfies $\Pr\{|J\cap j_i| > I^\prime\} = \delta^\prime$, then with probability $1-\delta^\prime$, $\Delta_{2,f_\mathrm{s}}=\sqrt{I^\prime}$.
Here, applying Hoeffding's inequality to $I$ i.i.d. Bernoulli random variables with a parameter $p$ yields that 
\begin{align*}
    \Pr\{|J\cap j_i| > I^\prime\} &\leq \Pr\{|J\cap j_i| \geq I^\prime\}\\
    &\leq \exp \left(-2I (p-\frac{I^\prime}{I})^{2}\right),
\end{align*}
for $I^\prime\geq Ip$.
Rearranging the equality $\delta^\prime = \exp \left(-2I (p-\frac{I^\prime}{I})^{2}\right)$ implies that for 
$I^\prime$ s.t. $I^\prime = Ip+\sqrt{\frac{I}{2} \log 1/\delta^\prime}$, it holds that with probability $1-\delta^\prime$, $\Delta_{2,f_\mathrm{s}}=\sqrt{I^\prime}$.
\end{proof}

\subsection{Proof of Theorem 4}
\begin{proof}
As argued in the main paper, $\Delta_{2, f_{\mathrm{fs}}}^{2} \leq \sigma_{\max}^2 (B) I$.
Furthermore, from Theorem~3, we know  
\begin{align*}
   \Pr \left\{\sigma^{2}_{\max }(B) \geq(1+\varepsilon) p \sigma^{2}_{\max}(A)\right\} 
    \leq 2\mathrm{srank}(A)\left(\frac{\mathrm{e}^{\varepsilon}}{(1+\varepsilon)^{1+\varepsilon}}\right)^{p\sigma^{2}_{\max }(A) / L} .
\end{align*}
By using the fact that $\sigma_{\max }(A) = 1$ and letting $\alpha = \sqrt{(1+\epsilon)p}$, we have
\begin{align*}
   \Pr \left\{\sigma^{2}_{\max }(B) \geq \alpha^2\right\} 
    \leq 2\mathrm{srank}(A)\left(\frac{\mathrm{e}^{\alpha^2/p-1}}{(\alpha^2/p)^{\alpha^2/p}}\right)^{p / L} .
\end{align*}
Thus, it holds that $\Pr \{\sigma_{\max }(B) \leq \alpha\} \geq 1-\delta^\prime$, where $\delta^\prime = 2\mathrm{srank}(A)\left(\frac{\mathrm{e}^{\alpha^2/p-1}}{(\alpha^2/p)^{\alpha^2/p}}\right)^{p / L}$.
Therefore, we have $\Pr\{\Delta_{2, f_{\mathrm{fs}}} \leq \alpha \sqrt{I}\} \geq 1-\delta^\prime$.
\end{proof}

\subsection{Proof of Corollary 2}
\begin{proof}
Let $M^\prime:\mathbb{R}^T\to\mathbb{R}^T$ be Algorithm~1 and
$M:\mathbb{R}^T\to\mathbb{R}^{T^\prime}$ be Algorithm~1 without the interpolation in the last step.
We define the interpolation map given $j\in\mathcal{J}$ as $\mathrm{Interpolate}_j:\mathbb{R}^{|j|} \to \mathbb{R}^T$.
Note that this map is bijective.

Let $S^{\prime} \subseteq \mathrm{range}(M^{\prime})$ be arbitrary.
Also, we denote the corresponding set by the inverse of $\mathrm{Interpolate}_j$ for $M$ as $S\subseteq \mathrm{range}(M)$.
Furthermore, we let $F\subseteq \mathcal{J}$ be the failure indices set. 
More formally, $F = \{j\in\mathcal{J}: \sigma_{\max}(B) > \alpha\}$, where $B = \mathrm{diag}(\delta_1, \ldots, \delta_T)A$.
Note that under $j\in F$, the sensitivity is $\sqrt{I}$, but under $j\in F^\complement$, it is $\alpha\sqrt{I}$.
Then, it holds that 
\begin{align*}
    \Pr \{M^{\prime}(x) \in S^{\prime}\} &= \sum_{j\in \mathcal{J}} \Pr\{M^{\prime}(x) \in S^{\prime}, J=j\}\\
    &= \sum_{j\in \mathcal{J}} \Pr\{M(x) \in S, J=j\}\\
    &= \sum_{j\in F} \Pr\{M(x) \in S| J=j\}\Pr\{J=j\} + \sum_{j\in F^{\complement}} \Pr\{M(x) \in S|J=j\}\Pr\{J=j\}\\
    &\leq\sum_{j\in F} (\exp(\epsilon/\alpha)\Pr\{M(x^{\prime}) \in S| J=j\}+\delta)\Pr\{J=j\} \\
    &\quad + \sum_{j\in F^{\complement}} (\exp(\epsilon)\Pr\{M(x^{\prime}) \in S|J=j\}+\delta)\Pr\{J=j\}\\
    &=\exp(\epsilon/\alpha)\sum_{j\in F} \Pr\{M(x^{\prime}) \in S| J=j\}\Pr\{J=j\}+\delta\Pr\{J\in F\}\\
    &\quad + \exp(\epsilon)\sum_{j\in F^{\complement}} \Pr\{M(x^{\prime}) \in S|J=j\}\Pr\{J=j\}+\delta\Pr\{J\in F^{\complement}\}\\
    &=\exp(\epsilon)\sum_{j\in\mathcal{J}} \Pr\{M(x^{\prime}) \in S,J=j\}
    +\delta+(\exp(\epsilon/\alpha)-\exp(\epsilon))\sum_{j\in F} \Pr\{M(x^{\prime}) \in S| J=j\}\Pr\{J=j\}\\
    &=\exp(\epsilon)\sum_{j\in\mathcal{J}} \Pr\{M^{\prime}(x^{\prime}) \in S^{\prime},J=j\}
    +\delta+(\exp(\epsilon/\alpha)-\exp(\epsilon))\sum_{j\in F} \Pr\{M(x^{\prime}) \in S| J=j\}\Pr\{J=j\}\\
    &\leq\exp(\epsilon)\Pr\{M^\prime(x^{\prime})\in S^{\prime}\}+\delta+(\exp(\epsilon/\alpha)-\exp(\epsilon))\Pr\{J\in F\}\\
    &\leq\exp(\epsilon)\Pr\{M^\prime(x^{\prime}) \in S^{\prime}\}+\delta+\delta^{\prime}(\exp(\epsilon/\alpha)-\exp(\epsilon))
\end{align*}
where the first inequality follows from the use of the Gaussian mechanism, 
the second inequality holds due to $\Pr[M(x^{\prime}) \in S| J=j] \leq 1$, and 
the last inequality follows by Theorem~4.
\end{proof}

\subsection{Proof of Corollary 3}
\begin{proof}
If $A$ is an identity matrix, we are able to use the sensitivity bound in Theorem~2; thus, $\delta^\prime$ is the one in Theorem~2 with $I^\prime =  \left\lceil \alpha^2 I \right\rceil$.
Note that $I^\prime \in \mathbb{N}$.
Let $F = \{j\in\mathcal{J} : |j\cap j_i| > I^\prime\}$. Then, under $j \in F^\complement$, the sensitivity is $\sqrt{I^\prime}$.
Finally, replacing $\alpha$ with $\sqrt{I^\prime/I}$ and proceeding the same transformation as the proof of Corollary~2 yield the final claim, where we use Theorem~2 to bound $\Pr\{J\in F\}$ in the final step.
\end{proof}

\subsection{Proof of Proposition 1}
\begin{proof}
By the assumption that $\sup_i \sum_t d_{i,t} = cI$ for some $c>1$, the sensitivity becomes 
$\sqrt{cI}$ under $j\in F$ and $\alpha\sqrt{cI}$ under $j\in F^\complement$, where $F$ is the failure indices set defined in the proof of Corollary~2.
Therefore, if we replace $\epsilon$ in Algorithm~1 with $\sqrt{c}\epsilon$, then we can proceed the identical transformation as the proof of Corollary~2.
Thus, we have the privacy guarantee that replaces $\epsilon$ in Corollary~2 with $\sqrt{c}\epsilon$ for the case where $\sup_i \sum_t d_{i,t} = cI$ for some $c>1$.
\end{proof}

\section{Details of Experiments}
\subsection{Instantiation of Our Algorithm}
We instantiate the filter with the gaussian filter.
More formally, we define the filtering vector $h$ that constitutes $A$ by 
\begin{align*}
    \hat{h}_t &= \exp\left(-\frac{1}{2}\left(\frac{T/2-|t-T/2|}{\sigma_g}\right)^2\right)\quad \forall t\in [T]\\
    h &= \hat{h} / \|\hat{h}\|_1.
\end{align*}
We set $\sigma_g = 10$ for the real datasets and $\sigma_g=10 / (1-\log(f)/4)$ for the synthetic dataset.

We implement the interpolation in Algorithm~1 by SciPy's \citep{virtanen_scipy_2020} \texttt{scipy.interpolate.interp1d} method with \texttt{kind} parameter set to \texttt{linear}. 
There's a case that $J$ in Algorithm~1 does not contain the first and last indices; thus, we need to extrapolate the signal.
We deal with it by copying the nearest value.
For example, if $\min J = 2$, we publish $\hat{z} = (z_2,z_2,\ldots)^\top$.

\subsection{Why We Fix $I$ for Synthetic Data}
In our experiment with Synth data, we fix $I$ for any (relative) sampling frequency $f$ even though different sampling frequencies $f$ yield different numbers of time steps $T$:
the larger $f$ becomes, the larger $T$ becomes. However, for e.g., the PeMS data (traffic flow), $I$ stays constant and does not scale with $f$. For example, if we assume a commuter passes a given sensor at most once every hour, then as long as the sampling frequency remains above once an hour, $I$ will stay fixed regardless of $T$. 
In other words, we can consider the number of time steps, $T$, as $T = (\text{Duration}) \times (\text{Sampling Frequency})$, 
and (for a high enough sampling frequency $f$) $I$ only depends on signal duration. 
In fact, in the instance that sampling frequency descends below once per hour, sensitivity still remains unchanged given the above assumption. For instance, if the sampling frequency drops to once every two hours, an individual may participate \emph{twice} in a time-step. While this may seem to invalidate our assumption of single participation per time-step ($M = 1$), the sensitivity stays constant since the product $MI$ (number of participation by an individual in a fixed duration of the time-series) remains constant. 

On the other hand, for real signals, the sampling frequency tends to be fixed; thus, $I$ scales with $T$ since larger $T$ means longer observation duration.
Therefore, it is indeed necessary to reduce the sensitivity scaling with $T$ or $I$ as an analyst wants to use a signal observed for a longer period with a given frequency $f$.

\subsection{Outputs Operating on the Noiseless Synth Signal}
\begin{figure}[t]
\centering
\includegraphics[width=\textwidth]{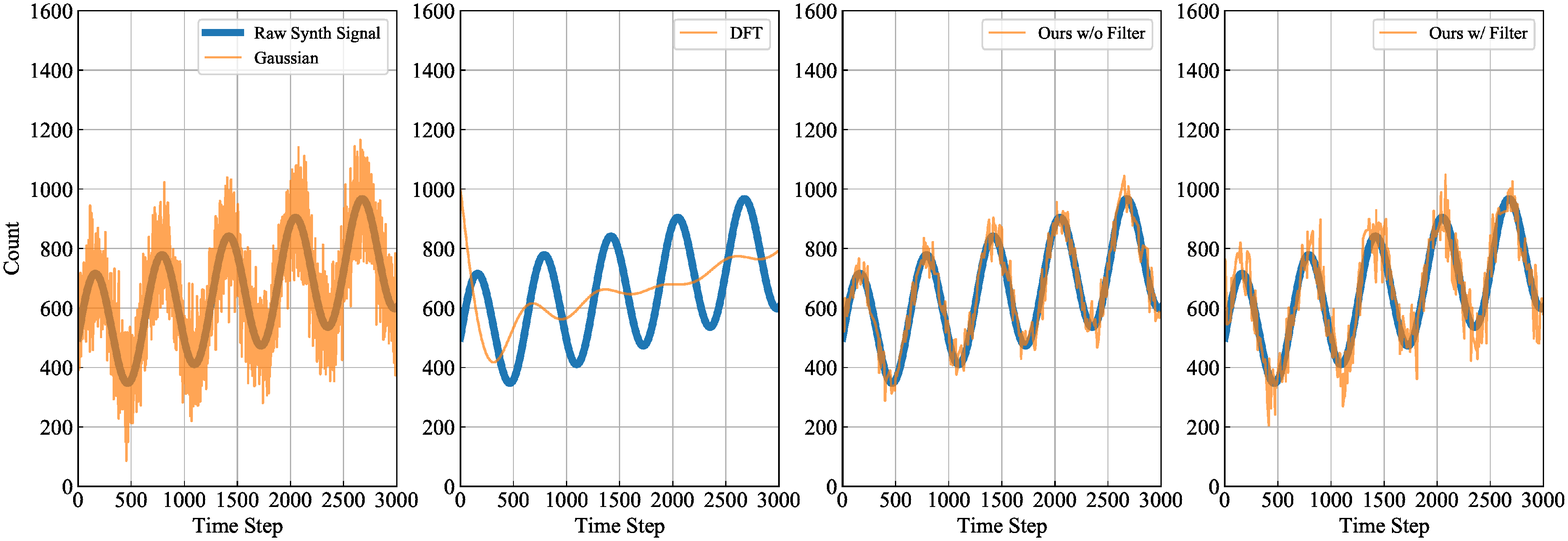}
\caption{
Two baseline mechanisms (left) and our mechanism with and without filtering (right), each operating on Synth time-series without observation noise. We plot the noiseless Synth time-series in blue for reference. 
Output by our algorithm without filtering concentrates around raw signal the most, indicating superiority of our algorithm with subsampling alone for smooth time-series signals.
}
\label{fig:synth_signal_noiseless}
\end{figure}
Figure~\ref{fig:synth_signal_noiseless} shows in orange examples of outputs of four mechanisms: the two baseline mechanisms and our mechanism with and without filtering, each taking the Synth signal without observation noise as input.
For reference the noiseless Synth signal is plotted in blue. 
Similar to the case where each mechanism takes the noisy Synth signal as input, outputs from our mechanism with and without filtering concentrate around the raw signal.
Comparing the outputs by our mechanism with and without filtering, we see that subsampling alone helps produce less noisy output and capture underlying trends better.
This supports the fact that the sensitivity reduction is larger for subsampling without filtering in general.

\section{Miscellany}
\subsection{Relationship between $\alpha$ and $\delta^\prime$ in Theorem~4}
\begin{figure}[t]
\centering
\includegraphics[width=0.9\textwidth]{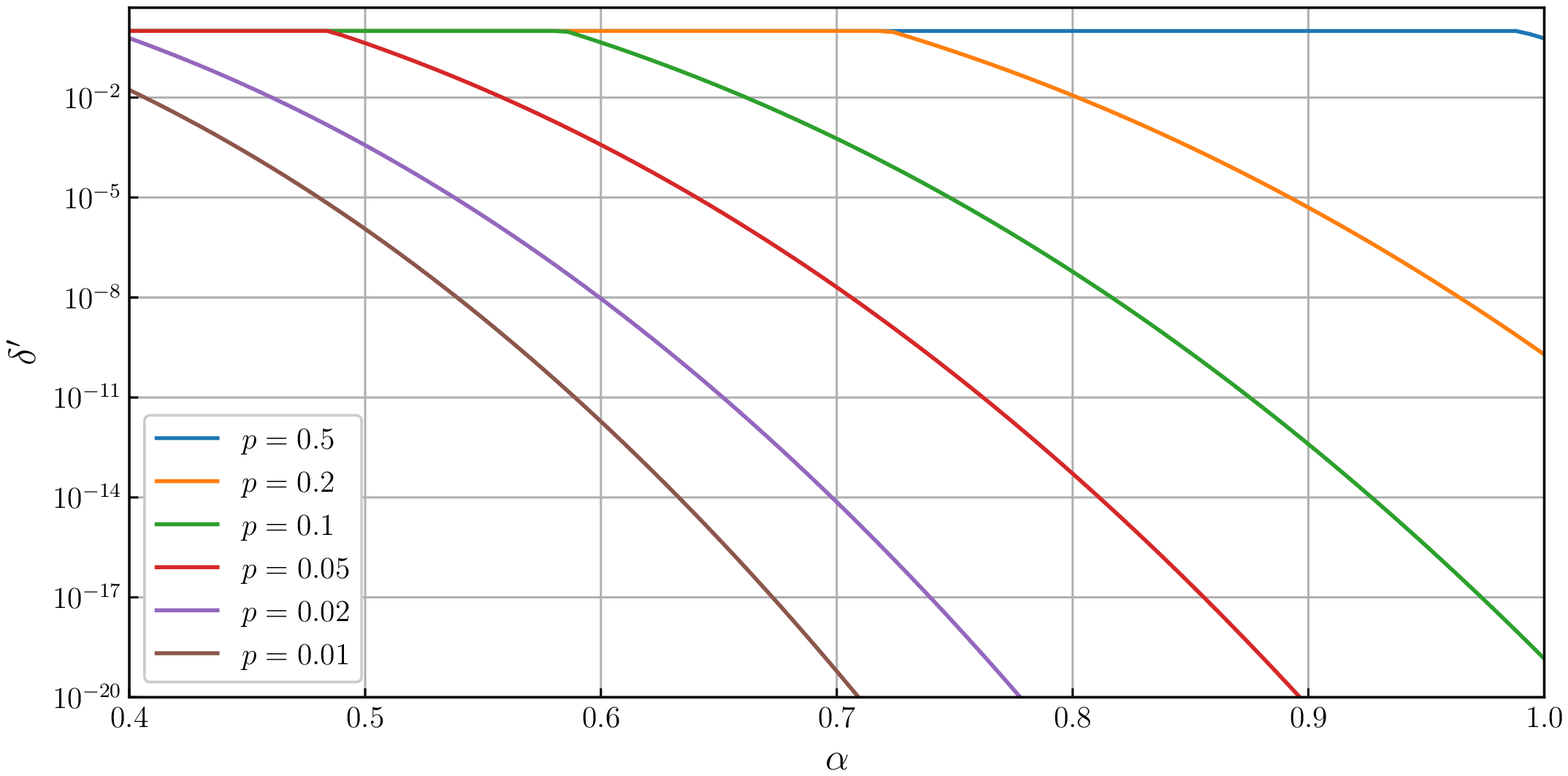}
\caption{
Tail probability, $\delta^\prime$, of $\Delta_{2, f_{\mathrm{fs}}}$ exceeding $\alpha \sqrt{I}$ given $\alpha$ in Theorem~4 for varying subsampling parameter $p$.
We observe that decrease rate of $\delta^\prime$ is at least exponentially fast. 
We also see that as $p$ becomes smaller we get more sensitivity reduction.
}
\label{fig:alpha_vs_delta}
\end{figure}
Figure~\ref{fig:alpha_vs_delta} illustrates the tail probability $\delta^\prime$ of $\Delta_{2, f_{\mathrm{fs}}}$ exceeding $\alpha \sqrt{I}$ given $\alpha \in [0,1]$,
i.e., $\delta^\prime = \Pr\{\Delta_{2, f_{\mathrm{fs}}} > \alpha \sqrt{I}\}$.
We set other parameters as $T=10000$ and $\sigma_g=10$, yielding $\mathrm{srank}(A) \approx 280$ and $L \approx 0.028$ in Theorem~4.\\
As expected, we get more sensitivity reduction as subsampling parameter $p$ gets smaller.
More importantly, we observe that $\delta^\prime$ decreases at least exponentially.
We believe the reduction is still notable though it is smaller for filtering and subsampling compared with subsampling alone.

\bibliography{supplement}